\documentclass[conference]{IEEEtran}
\IEEEoverridecommandlockouts
\usepackage{cite}
\usepackage{amsmath,amssymb,amsfonts}
\usepackage{algorithmic}
\usepackage{graphicx}
\usepackage{textcomp}
\usepackage{xcolor}
\def\BibTeX{{\rm B\kern-.05em{\sc i\kern-.025em b}\kern-.08em
    T\kern-.1667em\lower.7ex\hbox{E}\kern-.125emX}}
\usepackage{listings, multicol}
\definecolor{delim}{RGB}{20,105,176}
\definecolor{numb}{RGB}{106, 109, 32}
\definecolor{string}{rgb}{0.64,0.08,0.08}
\lstdefinelanguage{json}{
    showspaces=false,
    showtabs=false,
    breaklines=true,
    breakatwhitespace=true,
    postbreak=\raisebox{0ex}[0ex][0ex]{\ensuremath{\color{gray}\hookrightarrow\space}},
    basicstyle=\ttfamily\small,
    upquote=true,
    morestring=[b]",
    stringstyle=\color{string},
    literate=
     *{0}{{{\color{numb}0}}}{1}
      {1}{{{\color{numb}1}}}{1}
      {2}{{{\color{numb}2}}}{1}
      {3}{{{\color{numb}3}}}{1}
      {4}{{{\color{numb}4}}}{1}
      {5}{{{\color{numb}5}}}{1}
      {6}{{{\color{numb}6}}}{1}
      {7}{{{\color{numb}7}}}{1}
      {8}{{{\color{numb}8}}}{1}
      {9}{{{\color{numb}9}}}{1}
      {\{}{{{\color{delim}{\{}}}}{1}
      {\}}{{{\color{delim}{\}}}}}{1}
      {[}{{{\color{delim}{[}}}}{1}
      {]}{{{\color{delim}{]}}}}{1},
}
\lstdefinelanguage{SQL}{
   breaklines=true,
   postbreak=\mbox{\textcolor{gray}{$\hookrightarrow$}\space},
   breakatwhitespace=true,
   keywords={TYPE, DATASET, CREATE, FEED, WITH, START, STOP, TO, CHANNEL, BROKER, INDEX, FUNCTION, LET, GROUP, BY, GROUP, AS, SELECT, WHERE, FROM, ORDER, DESC, LIMIT, REPETITIVE, SUBSCRIBE, ON, AT, USING, AND, PERIOD, ACTIVE, PRIMARY, KEY, CONTINUOUS, CONNECT, APPLY, PUSH, USE, UNNEST}
}
\makeatletter
\newcommand\notsotiny{\@setfontsize\notsotiny\@vipt\@viipt}
\makeatother
\begin{document}
\title{Bridging BAD Islands: \\ Declarative Data Sharing at Scale}


\author{\IEEEauthorblockN{Xikui Wang, Michael J. Carey}
\IEEEauthorblockA{\textit{Donald Bren School of Information and Computer Sciences} \\
\textit{University of California Irvine}\\
Irvine, United States\\
\{xikuiw, mjcarey\}@ics.uci.edu}
\and
\IEEEauthorblockN{Vassilis J. Tsotras}
\IEEEauthorblockA{\textit{Department of Computer Science and Engineering} \\
\textit{University of California Riverside}\\
Riverside, United States\\
tsotras@cs.ucr.edu}
}

\IEEEoverridecommandlockouts
\IEEEpubid{\makebox[\columnwidth]{978-1-7281-6251-5/20/\$31.00~\copyright2020 IEEE \hfill} \hspace{\columnsep}\makebox[\columnwidth]{ }}

\maketitle

\IEEEpubidadjcol

\begin{abstract}
In many Big Data applications today, information needs to be actively shared between systems managed by different organizations. To enable sharing Big Data at scale, developers would have to create dedicated server programs and glue together multiple Big Data systems for scalability. Developing and managing such glued data sharing services requires a significant amount of work from developers. In our prior work, we developed a Big Active Data (BAD) system for enabling Big Data subscriptions and analytics with millions of subscribers. Based on that, we introduce a new mechanism for enabling the sharing of Big Data at scale \textit{declaratively} so that developers can easily create and provide data sharing services using declarative statements and can benefit from an underlying scalable infrastructure. We show our implementation on top of the BAD system, explain the data sharing data flow among multiple systems, and present a prototype system with experimental results.
\end{abstract}


\section{Introduction}
\label{sec:bad_island_overview}
Advances in information technology have created large collections of data~\cite{big_data_computing}. Such large volumes of data - Big Data - also come with big challenges. In order to transmit, process, and persist Big Data, researchers and experts from academia and industry have developed a plethora of systems~\cite{shvachko2010hadoop, pig:paper, HivePaper, Spark}. However, most of them are passive in nature - passively answering users' requests to process and return data rather than actively processing and delivering data of interest to users. In many applications, users not only want to analyze data, but also to subscribe to and actively receive data of interest. Their interests may include the data's content as well as its relationships to other data. For example, in-field officers may want to \textit{receive \textbf{nearby} threatening tweets \textbf{whenever they are posted}}. There can be millions of users having similar requests. 
We refer to the enabling of Big Data subscriptions and analytics as \textit{Big Active Data} (BAD). 
Traditional pub/sub systems~\cite{many_faces} often lack the capability of data processing and handling complex subscription requests that involve data's relationships (\textit{such as send me tweets \textbf{near my current location}}). More recent stream processing engines~\cite{spark_streaming_paper,flink} usually don't persist data for historical data analytics (\textit{such as show me the average threatening rating of tweets in the \textbf{past five months} grouped by their location}). In order to accommodate BAD challenges, we have created a BAD system that supports Big Data subscriptions and analytics at scale~\cite{breakingbad,bad_to_the_bone,jacobs2017bad,xikui_thesis,wang2020subscribing}.

In a big BAD world, the data to be analyzed and delivered often needs to be processed and enriched with additional information so that interested users can obtain more insights from the data. Such additional information may be managed by different organizations.
Developers often need to share data between different systems for supporting BAD applications (e.g., threatening tweets detected at the Department of Homeland Security need to be shared with local police departments). Data sharing can be difficult, besides the ethical and legal issues, because of the challenges in management, interoperability, security, and infrastructure~\cite{study_on_data_sharing}. Researchers from academia have developed projects that unify institutional repositories from different organizations for sharing research datasets~\cite{dataone,disc_uk}. Companies have also created platforms based on Big Data projects to improve business efficiency and consolidate resources for better services~\cite{data_sharing_in_europe}. Nevertheless, providing efficient, reliable, and scalable data sharing services require dedicated infrastructures and collaborative efforts from developers and organizations. In this work, we focus on enabling the active sharing of Big Data declaratively in a BAD world. In particular, we characterize a BAD world as a group of BAD islands, where each organization runs an independent BAD system as an island. We discuss how to ``bridge'' different BAD islands using scalable data sharing services without additional programming from developers. 


\section{Big Active Data in a Nutshell}
\label{sec:bad_in_a_nutshell}
\begin{figure}[h]
    \centering
    \includegraphics[width=.48\textwidth]{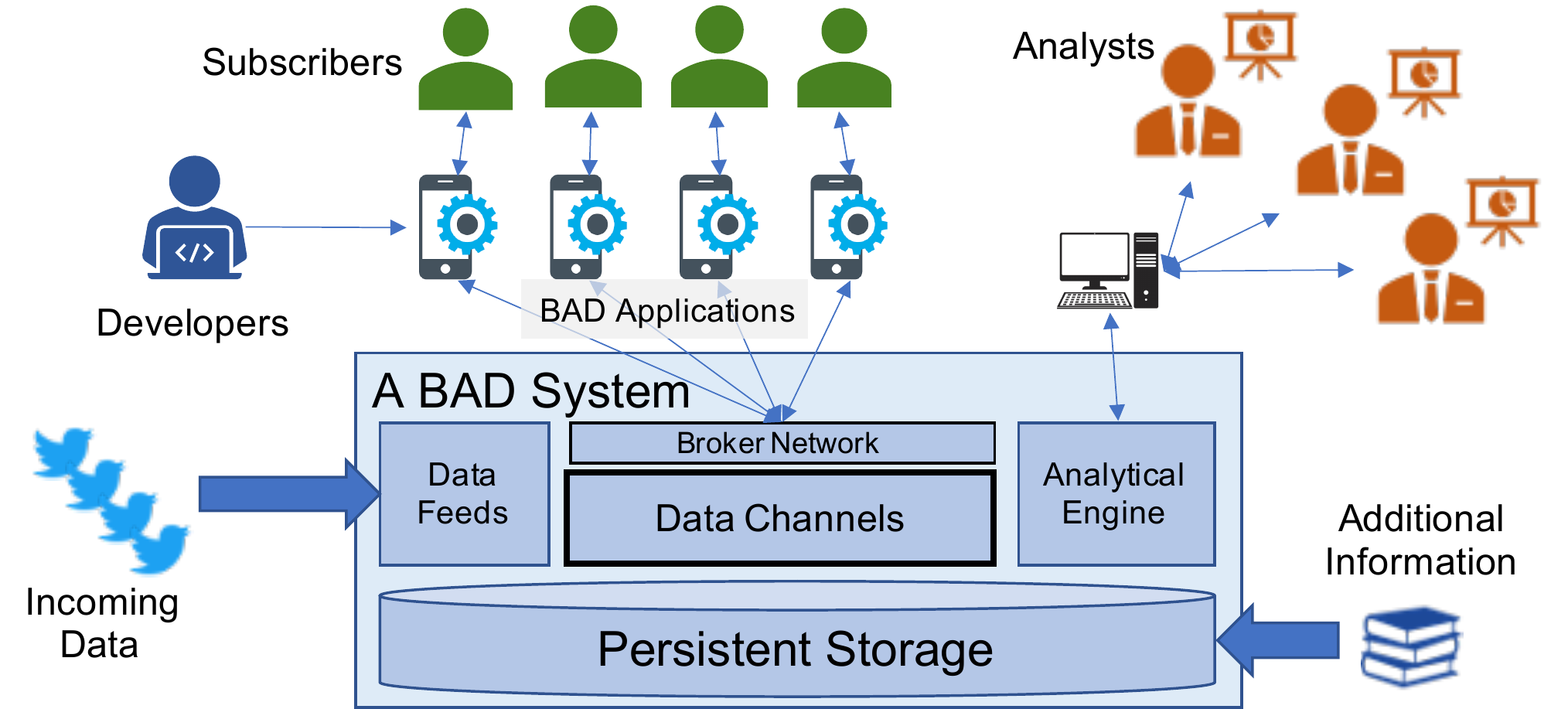}
    \caption{An overview of the BAD system}
    \vspace{-3mm}
    \label{fig:bad_overview}
\end{figure}

Our BAD system has been built as an extension of Apache AsterixDB, a Big Data Management System (BDMS) that provides distributed data management for large-scale, semi-structured data~\cite{asterixdb14}.
The BAD system \cite{bad_to_the_bone,wang2020subscribing} can enable millions of users to subscribe to data of interest and receive updates continuously, and it also supports Big Data analytics with a declarative language, SQL++ (a SQL-inspired query language for semi-structured data)~\cite{sql++don}. An overview of the BAD system is shown in Figure~\ref{fig:bad_overview}. Due to space limits, here we focus on two key components: Data Feeds and Data Channels. 
For more details about our project we refer to \cite{breakingbad,bad_to_the_bone,jacobs2017bad,xikui_thesis,wang2020subscribing}.

\vspace{-4mm}
\subsection{Data Feeds}
Data feeds help the BAD system to ingest rapidly incoming data from various external data sources in different formats. Users can create a data feed using SQL++ statements. As an example, in Figure~\ref{ddl:ing_feed}, we define a data type \textit{Tweet} to describe the incoming data's minimum required attributes and an active dataset \textit{Tweets} to persist the incoming data. 
Active datasets, different from normal datasets, enable continuous query semantics~\cite{tapestry} in channels (discussed next)~\cite{xikui_thesis,bad_to_the_bone}.
Here we create a \textit{TweetFeed} using a socket adapter and specify the incoming data's format as JSON. This allows the BAD system to use a socket server to intake incoming JSON data. The \textit{TweetFeed} is connected to the dataset \textit{Tweets} so that the ingested data can be persisted in storage (partitioned across all nodes of a cluster) directly for later use. There are two types of data feeds: static feeds, which maximize ingestion throughput, and dynamic feeds, which allow users to enrich incoming data using user-defined functions (UDFs)~\cite{DBLP:conf/sigmod/AlkowaileetA0LR18,DBLP:journals/pvldb/WangC19,DBLP:conf/edbt/GroverC15,DBLP:conf/semco/AlkowaileetACLR18}.
\begin{figure}[h]
\notsotiny
\vspace{-8mm}
\begin{lstlisting}[
           language=SQL,
           basicstyle=\ttfamily,
           showstringspaces=false,
           commentstyle=\color{gray}
        ]    
CREATE TYPE Tweet AS { tid: bigint, uid: bigint, text: string };
CREATE ACTIVE DATASET Tweets(Tweet) PRIMARY KEY tid;
CREATE FEED TweetFeed WITH { 
  "type-name" : "TweetType", 
  "adapter-name": "socket_adapter", 
  "format" : "JSON", 
  "sockets": "FEED_HOST:FEED_PORT", 
  "address-type": "IP", 
  "dynamic": false };
CONNECT FEED TweetFeed TO DATASET Tweets;
START FEED TweetFeed;
\end{lstlisting}
\vspace{-2mm}
\caption{A sample data feed connected to an active dataset}
\vspace{-4mm}
\label{ddl:ing_feed}
\end{figure}

\subsection{Data Channels}
Data channels allow developers to activate parameterized queries as services for millions of users to subscribe to and continuously receive their data of interest. 
When creating a channel, developers can construct a \textit{channel query} to describe the data of interest for subscribers and specify a \textit{channel period} to indicate how often should the channel query be evaluated for subscribed users. All subscriptions of a channel are evaluated together to allow the system to exploit shared computations among them (e.g., many subscribers could be interested in tweets from Orange County), and increasing the channel period could lead to a bigger batch size and thus allow computing complex data of interest for more subscribers at scale.
For example, we can create a \textit{NearbyThreateningTweets} channel, as shown in Figure~\ref{ddl:channel}, to allow in-field officers to subscribe to nearby threatening tweets. 
In the channel query, we use the \textit{is\_new} function to look for \textbf{new} threatening tweets near a subscribed officer's location and return those tweets to subscribers every 10 seconds.\footnote{One could also apply the \textit{is\_new} function on OfficerLocations to look for nearby threatening tweets only for officers actively updating their locations. Interested readers may refer to \cite{wang2020subscribing} for more continuous channel examples.} The active dataset \textit{Tweets} provides continuous query semantics to make sure every qualified new tweet will be delivered to subscribed officers. The threatening tweets for subscribers are sent to brokers registered as HTTP endpoints in the BAD system. A user can subscribe to a data channel on a broker and thus receive updates from it. As shown in Figure~\ref{ddl:broker_subscription}, we can register a broker and make two separate subscriptions (on behalf of in-field officers) on this broker so that the threatening tweets near these two in-field officers are sent to this broker and then delivered to them. Data channels provide two modes for delivering data: \textit{push} and \textit{pull}. In the push mode, the data of interest is pushed to brokers directly. In the pull mode, a broker having new data of interest for its subscribers will receive a notification from the channel, and then the broker can pull that data from BAD storage later.
\begin{figure}[h]
\notsotiny
\vspace{-5mm}
\begin{lstlisting}[
           language=SQL,
           basicstyle=\ttfamily,
           showstringspaces=false,
           commentstyle=\color{gray}
        ]
// Similar to TweetFeed and Tweets, we have a LocationFeed connected 
// to an OfficerLocations dataset to receive and store the live 
// location updates from in-field officers 
// CREATE TYPE OfficerLocation AS { oid: int, location: point };
// CREATE ACTIVE DATASET OfficerLocations(OfficerLocation) 
//   PRIMARY KEY oid;

CREATE CONTINUOUS CHANNEL NearbyThreateningTweets(oid) 
 PERIOD duration("PT10S") {
  SELECT t FROM OfficerLocations o, Tweets t
  WHERE spatial_distance(t.location, o.location) < 5 
    AND o.oid = oid AND t.threatening_rating > 0 AND is_new(t) };
\end{lstlisting}
\vspace{-2mm}
\caption{A sample continuous channel for nearby hateful tweets}
\vspace{-3mm}
\label{ddl:channel}
\end{figure}
\vspace{-3mm}

\begin{figure}[h]
\notsotiny
\begin{lstlisting}[
           language=SQL,
           basicstyle=\ttfamily,
           showstringspaces=false,
           commentstyle=\color{gray}
        ]    
CREATE BROKER BROKER_A AT "http://BROKER_A_HOST:BROKER_A_PORT/API";
SUBSCRIBE TO NearbyThreateningTweets("0907") ON BROKER_A;
SUBSCRIBE TO NearbyThreateningTweets("1226") ON BROKER_A;
\end{lstlisting}
\vspace{-2mm}
\caption{Registering a broker and making subscriptions}
\vspace{-3mm}
\label{ddl:broker_subscription}
\end{figure}

\section{BAD Islands}
\label{sec:three_islands_example}
In the following sections, we discuss how we can connect BAD systems managed by different organizations (islands) in a BAD world together to enable data sharing among them. We use a three-island example with the following organizations for illustration: the Department of Homeland Security, the Orange County Sheriff's Department, and the University of California-Irvine. Each organization hosts an independent BAD system and serves its own BAD users with localized information.

\subsection{BAD Island 1: Department of Homeland Security}
The Department of Homeland Security (DHS) is a federal agency responsible for ensuring public security. 
In our example, DHS has access to all tweets posted in the United States. These tweets cannot be shared with other organizations directly due to licensing and privacy concerns, except for the tweets that are related to potential threats. The BAD system at DHS needs to provide data analytics on collected tweets and serve tweets to its agents through data channels.

\begin{figure}[h]
    \centering
    \includegraphics[width=0.42\textwidth]{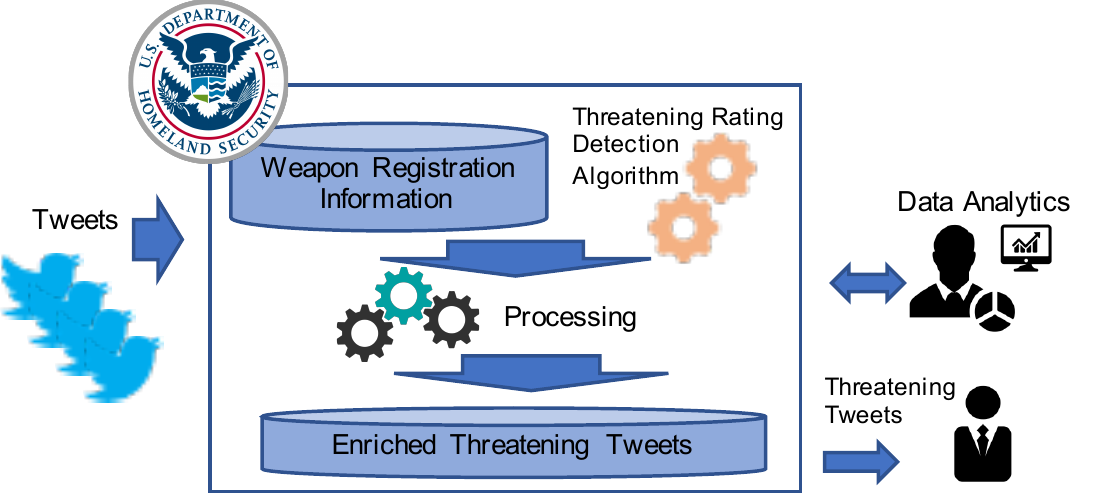}
    \vspace{-2mm}
    \caption{An overview of the DHS Island}
    \vspace{-4mm}
    \label{fig:dhs_overview}
\end{figure}

Since raw tweets from Twitter may not contain all necessary information, DHS might need to enrich them with other relevant data. As an example, DHS could collect weapon registration information for some sensitive twitter account holders and attach that to tweets to provide important additional information for interested subscribers. In addition, DHS could also utilize Machine Learning algorithms to estimate the threatening rating of the tweets' text and use that for later analysis. An overview of the DHS island is shown in Figure~\ref{fig:dhs_overview}.

\subsection{BAD Island 2: Orange County Sheriff's Department}
The Orange County Sheriff's Department (OCSD) is the local law enforcement agency that ensures safety and responds to potential crimes in Orange County, CA. In our use case, OCSD wants to monitor major local events and ensure the safety of the event and its participants.
In-field officers who patrol around the county, continuously report their locations back to OCSD so that OCSD can send them instructions based on their locations (e.g., when an emergency happens, send nearby officers for help).

\begin{figure}[h]
    \centering
    \vspace{-4mm}
    \includegraphics[width=0.42\textwidth]{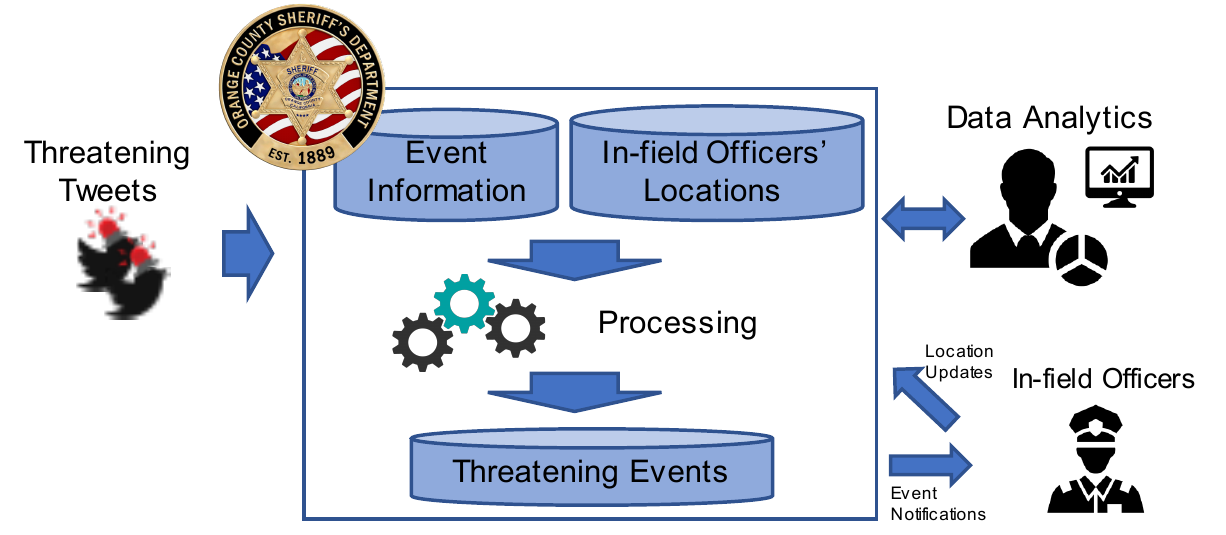}
    \vspace{-2mm}
    \caption{An overview of the OCSD island}
    \vspace{-4mm}
    \label{fig:ocsd_overview}
\end{figure}

To prevent potential threats to local events, OCSD would like to obtain the threatening tweets posted in Orange County. When a local threatening tweet is detected, OCSD can find important events close to the tweet and then notify the nearby in-field officers about the event and the tweet so they can further investigate it. Additionally, OCSD wants to support data analytics on data stored in the system. An overview of the OCSD island is shown in Figure~\ref{fig:ocsd_overview}.

\subsection{BAD Island 3: University of California-Irvine}
The University of California-Irvine (UCI) is a public university located in Irvine, a city in Orange County. 
The university often hosts various activities and events in different buildings on campus. To ensure students' and visitors' safety, the university has its own university police officers placed at various security stations on campus, and students/visitors can seek help from when an emergency happens. The buildings on campus have notice boards for showing important notifications and alerts. The university also has an alerting service - zotALERT - which delivers important messages to people (subscribers) on-campus through text messages and emails.

\begin{figure}[h]
    \centering
    \vspace{-4mm}
    \includegraphics[width=0.42\textwidth]{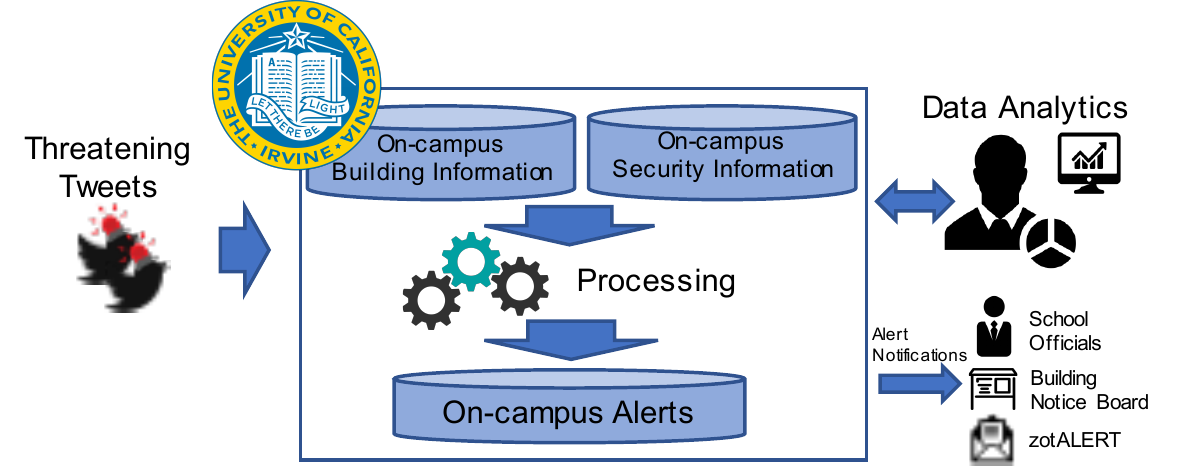}
    \vspace{-2mm}
    \caption{An overview of the UCI island}
    \vspace{-4mm}
    \label{fig:uci_overview}
\end{figure}

UCI would like to acquire the threatening tweets posted near the UCI campus and notify people in the buildings around those tweets to raise attention. An alert could include the information about nearby security stations for the tweet so that people in an emergency situation could quickly seek help. Data analytics on threatening tweets and other data in the system for school officials are also to be supported. An overview of the UCI island is shown in Figure~\ref{fig:uci_overview}.

\section{Island Hopping: Connecting BAD Islands}
\label{sec:conn_bad_islands}
In order to support the BAD services at OCSD and UCI described in Section~\ref{sec:three_islands_example}, we need to enable the sharing of threatening tweets detected at DHS with OCSD and UCI. These tweets can be combined with local information at Orange County and UCI, respectively, and then be used for creating localized notifications for subscribers on each island.
Below we consider three options for sharing threatening tweets among these islands, namely: (1) combining all islands together into one (\textit{a BAD Continent}), (2) creating direct connections between the individual islands as needed (\textit{BAD Ferries}) and (3) utilizing the channel idea to allow islands to subscribe to what they need from one another (\textit{BAD Bridges}). Below we discuss the three options in detail.

\subsection{Option 1: A BAD Continent}
\begin{figure}[h]
    \centering
    \vspace{-3mm}
    \includegraphics[width=0.42\textwidth]{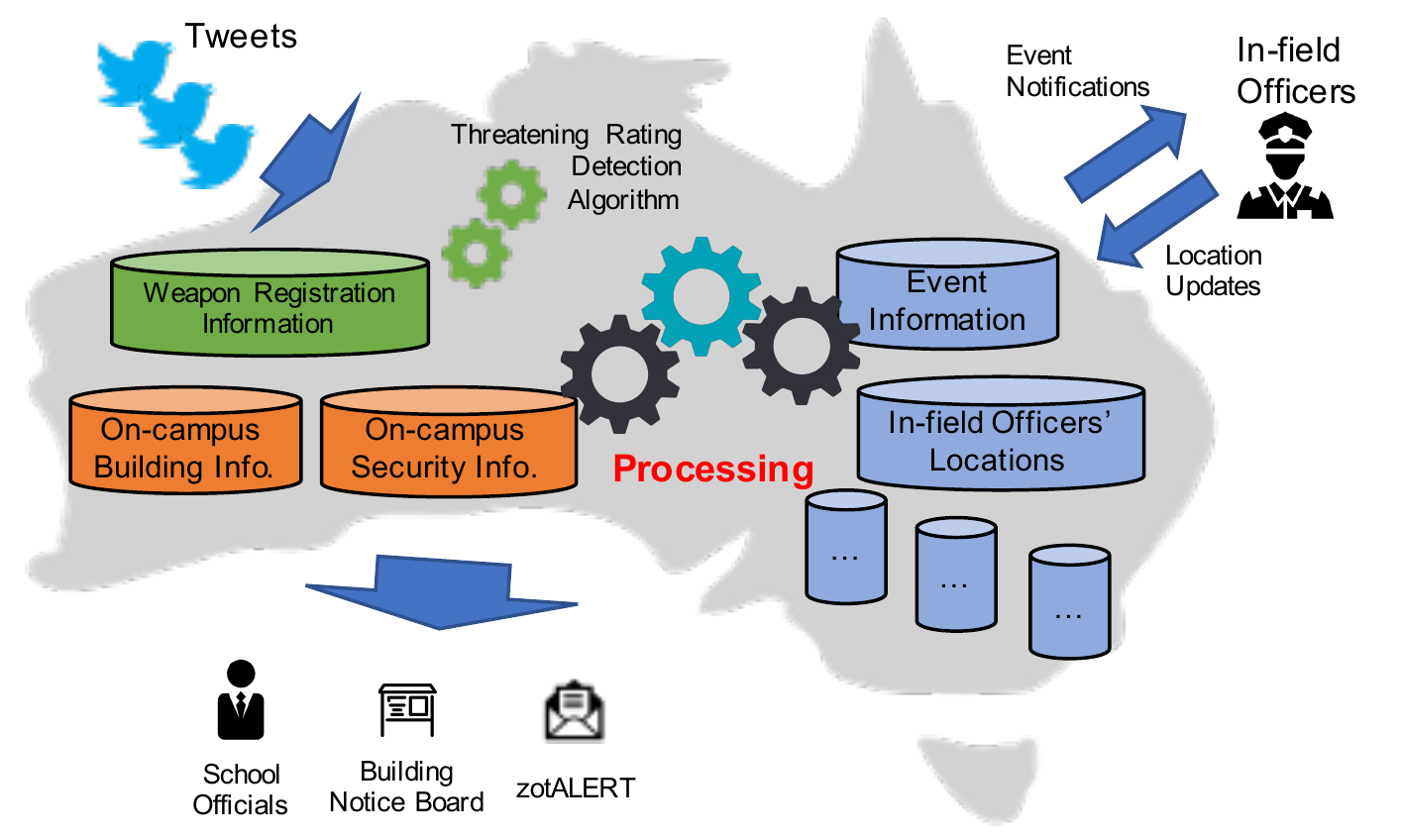}
    \vspace{-1mm}
    \caption{An illustration of a BAD continent}
    \vspace{-2mm}
    \label{fig:bad_continent}
\end{figure}

Instead of sharing threatening tweets between multiple BAD islands, one could create a big BAD island, namely a BAD continent, that holds not only the data at DHS but also the local data from OCSD and UCI, as shown in Figure~\ref{fig:bad_continent}. In this case, all services at OCSD and UCI could be integrated into this BAD continent, and all subscribers then would subscribe to this BAD continent directly. All information is now in the same system. Developers from different organizations could easily create BAD services without having to share data.

In principle, a one-for-all BAD continent could be easy to build, and it avoids the complexity of connecting different BAD islands.
Although the resulting BAD system could be scaled to support the volume of data and users from multiple organizations, such global integration would introduce significant management and administration overheads, especially for the service provider (DHS in this case). For the three-island example, not only would all local information (including local events, campus building layouts, etc.) need to be stored in the BAD continent, but all updates (location updates, event updates, etc.) would need to be forwarded to the system. Managing all local data at DHS could be very complex and would require sophisticated access control. When more organizations join, such a database would have to manage all kinds of additional local information while receiving updates from multiple parties; this system would quickly become impractical to maintain by one organization. Additionally, such global information sharing may not be permitted (by law) between different agencies in all cases.

\subsection{Option 2: BAD Ferries}
\begin{figure}[h]
    \centering
    \vspace{-3mm}
    \includegraphics[width=0.42\textwidth]{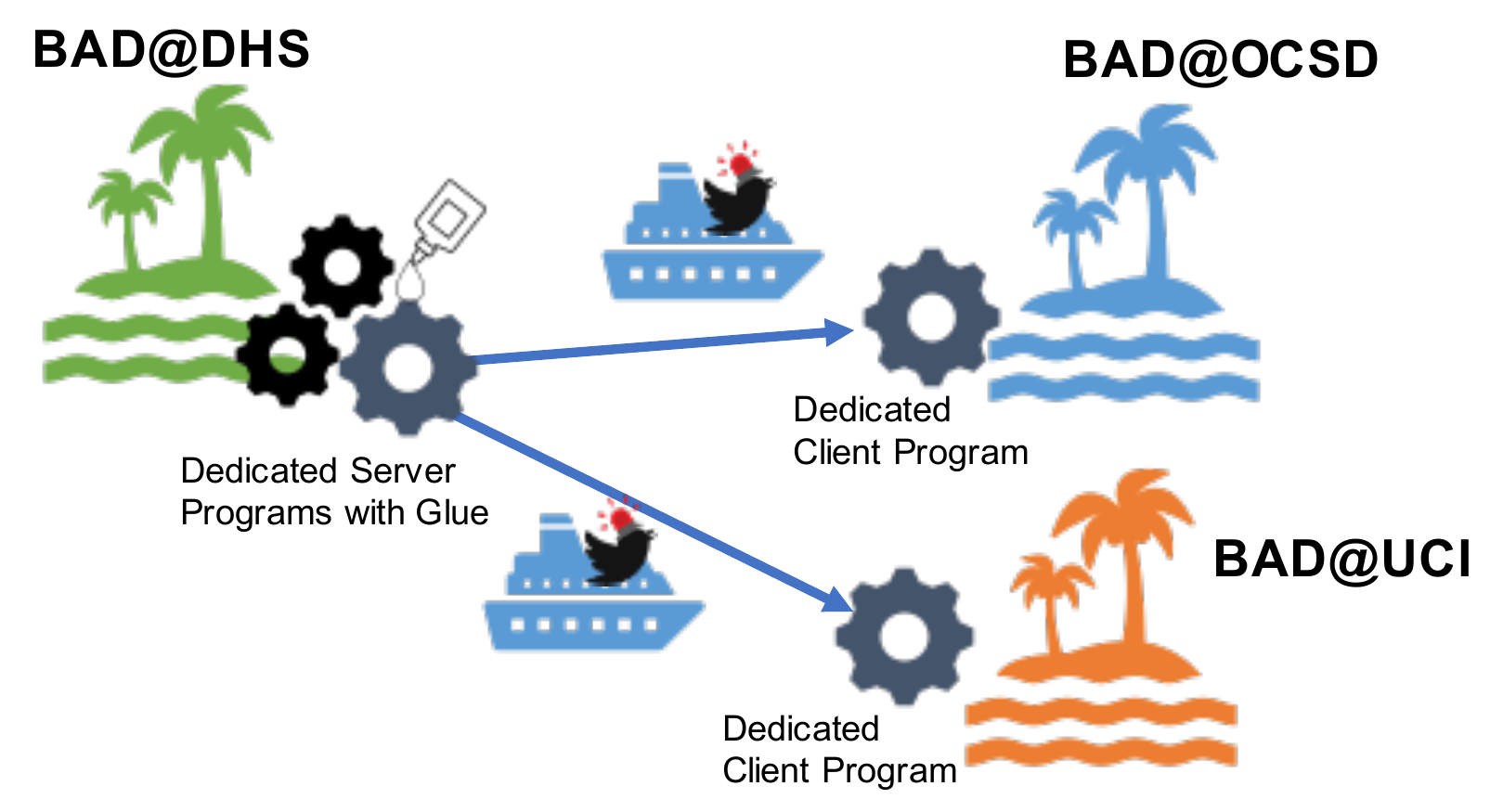}
    \vspace{-2mm}
    \caption{An illustration of BAD ferries}
    \vspace{-2mm}
    \label{fig:bad_ferries}
\end{figure}

A different way of supporting the required BAD services at OCSD and UCI, without combining everything together, would be to programmatically send the requested data from DHS to OCSD and UCI, as shown in Figure~\ref{fig:bad_ferries}. 
DHS could send the threatening tweets detected in Orange County and near UCI campus to OCSD and UCI, respectively, and OCSD BAD and UCI BAD could then combine those tweets with their local information to produce localized notifications for their subscribers.

In order to share the data cleanly and efficiently, DHS would need to create a dedicated server program that allows other organizations to access the shared data in DHS. Also, OCSD and UCI would need to develop corresponding client programs connected to the DHS server program and obtain shared data. Data exchanges between the server and clients could be frequent, and there could be many more clients who would like to access the shared data. Thus, the server program would need to be efficient, reliable, and scalable for handling a large number of clients and a large volume of data. Implementing and extending the server and client programs would require significant efforts from these organizations.

\subsection{Option 3: BAD Bridges}
\label{sec:bad_bridges}
\begin{figure}[h]
    \centering
    \vspace{-3mm}
    \includegraphics[width=0.42\textwidth]{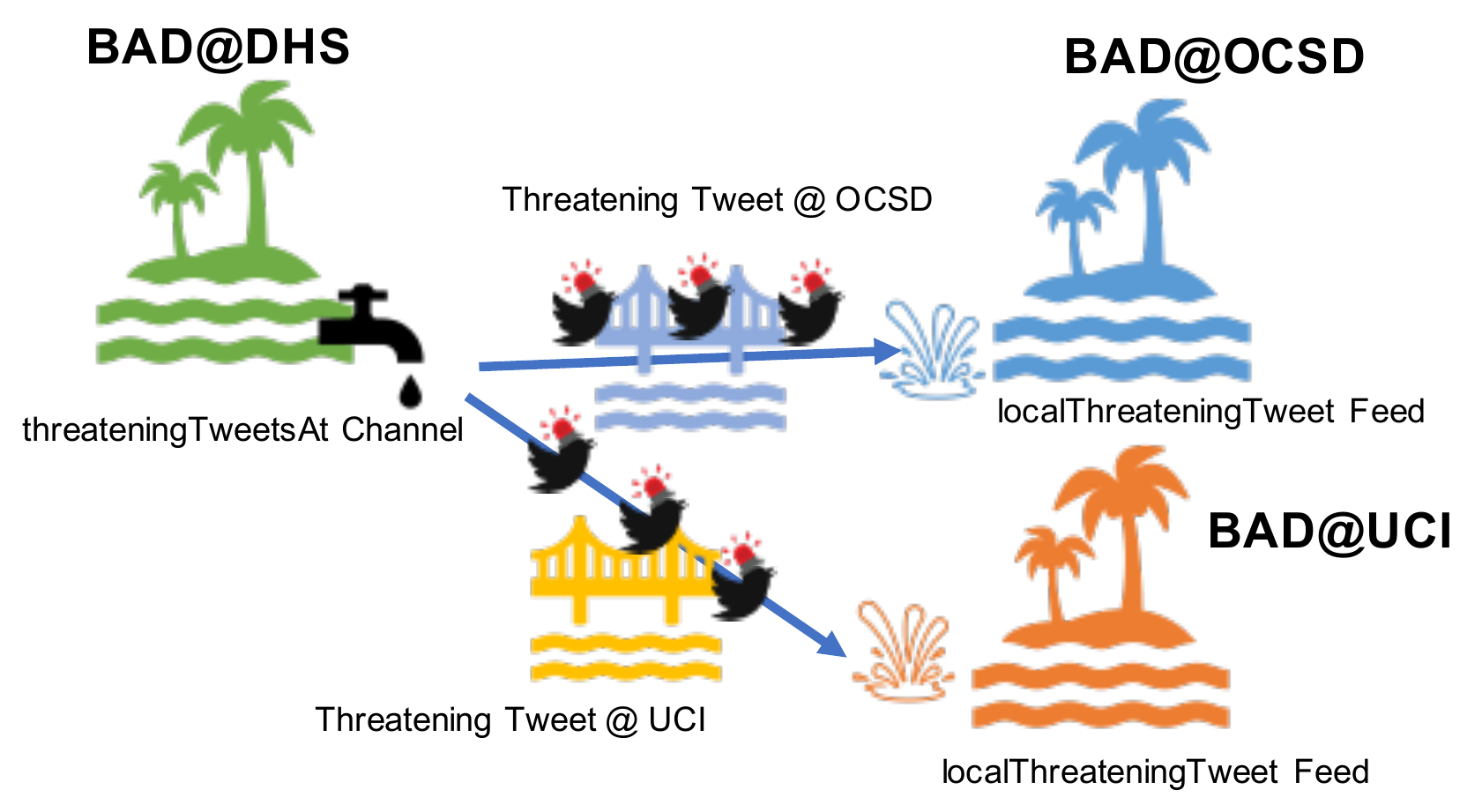}
    \vspace{-2mm}
    \caption{An illustration of BAD bridges}
    \vspace{-2mm}
    \label{fig:bad_bridges}
\end{figure}

An important observation is that this data exchange pattern,
where we have an island serving data and multiple islands constantly requesting data of interest, resonates well with the original BAD user model, where subscribers subscribe to data and constantly receive updates. Inspired by this, we could characterize a BAD island as being a BAD subscriber of another island and connect these islands using \textit{BAD bridges} built on data channels and data feeds to share data at scale, as shown in Figure~\ref{fig:bad_bridges}. 
One might characterize this architecture as: \textit{``One man's channel is another man's feed."}

Following our example, we could first create a data channel on DHS BAD, which serves threatening tweets by areas, namely via a threateningTweetsAt channel, and other islands interested in local threatening tweets from an area could then subscribe to this channel with the area name of interest. OCSD BAD, as a subscriber, can subscribe to this channel with the parameter ``OC'', and UCI BAD, as another subscriber, can also subscribe to this channel with the parameter ``UCI''. 
We could use a push channel to push threatening tweets to OCSD and UCI BAD so they can receive local threatening tweets from the channel at DHS directly, process them with local information, and then produce localized notifications to their own subscribers.

On OCSD and UCI BAD, we could utilize data feeds to receive threatening tweets detected by the threateningTweetsAt channel on DHS BAD. Taking OCSD BAD as an example, we could create an HTTP feed and connect it to a local OCSD dataset for persisting the threatening tweets. We could register the feed's HTTP address as a broker in DHS BAD and then subscribe to the threateningTweetsAt channel with the parameter ``OC''. With this feed, broker, and subscription, threatening tweets posted at Orange County and detected by DHS would then be sent to the feed's endpoint from the threateningTweetsAt channel. 
Similarly, we could repeat this process for other BAD systems to obtain threatening tweets from their areas of interest.
Since the BAD system is scalable and can support a large number of subscribers with a large volume of data, bridging BAD systems using data channels and feeds can be scaled out to support many more islands connecting to DHS. This allows developers to declaratively create data sharing services, without additional programming and gluing together multiple systems, as we will see next.

\section{Building BAD Bridges}
\label{sec:building_bad_bridges}
Given the advantages of the BAD Bridges approach, we now introduce \textit{BAD brokers} to further simplify and enhance data exchanges between BAD islands and \textit{BAD feeds} and thus help users create bridges and manage their life-cycles.

\subsection{BAD Brokers}
The broker sub-system in BAD manages the communication between the BAD system and its subscribers. A broker registers itself as an HTTP endpoint in the BAD system. Notifications containing data of interest produced by the BAD back-end are delivered to this broker endpoint and then disseminated to subscribers who subscribed on this broker. In order to allow general brokers to parse the incoming notifications, data channels produce notifications as JSON objects, and more complex data types supported in BAD in the AsterixDB Data Model (ADM) (such as datetimes, points, etc.) are encoded as strings, arrays, and other JSON data types. Since BAD islands are ``brokers'' that can also directly process ADM data, we can instead deliver their notifications as ADM records to maintain the richer data type information and avoid additional data encoding and decoding overheads.

To allow brokers to process ADM data and to become extensible for future use cases, we introduce a new notion of \textit{BAD brokers} and a simple new syntax for creating brokers in BAD, as shown in Figure~\ref{ddl:create_bad_broker}. Users can add an optional WITH statement for providing additional information about the broker. While we only support ``broker-type'' for now, this can be further extended to support other features in the future. When there is no WITH statement or when the broker-type is set to ``general", we create a general broker that takes JSON data. When the broker-type is set to ``BAD'', we create a BAD broker that takes ADM records. In general, a channel can have subscriptions from both types of brokers. In that case, channel executions will send JSON formatted data to the general brokers and ADM formatted data to the BAD brokers.
\begin{figure}[h]
\notsotiny
\vspace{-6mm}
\begin{lstlisting}[
           language=SQL,
           basicstyle=\ttfamily,
           showstringspaces=false,
           morekeywords={TYPE, DATASET, CREATE, FEED, WITH},
           commentstyle=\color{gray}
        ]
    CREATE BROKER BROKER_NAME AT "http://BROKER_HOST:PORT_NUM" WITH 
      { "broker-type" : "BAD" };
\end{lstlisting}
\vspace{-2mm}
\caption{Creating a BAD broker}
\vspace{-2mm}
\label{ddl:create_bad_broker}
\end{figure}

\vspace{-3mm}

\subsection{BAD Feeds}
Bridging from a BAD island \textit{A} to another BAD island \textit{B} and sharing data to island \textit{B} requires several steps: create a data feed on island \textit{B}; register the feed with island \textit{A} as a BAD broker; and create a subscription on island \textit{A} on the created broker. Also, removing the bridge between island \textit{A} and island \textit{B} requires unsubscribing from the channel and removing the BAD broker on island \textit{A}. In order to simplify the process of bridging BAD islands and help users manage the life-cycles of bridges, we also introduce the notion of \textit{BAD feeds}.

One can create a BAD feed on island \textit{B} and connect it to a channel on island \textit{A} using the statement in Figure~\ref{ddl:create_bad_feed}. 
Unlike regular data feeds, users would need to specify several additional configuration entries for connecting to a data channel on the other BAD island.
In particular, the ``bad-host'', ``bad-channel'', and ``bad-dataverse'' configuration parameters help the system locate the data channel on the other island, while ``bad-channel-parameters'' contains subscription parameters as a quote-escaped string for subscribing to the channel. When a channel takes multiple parameters, we use commas to separate them. If a data feed wants to subscribe to a channel with several different parameters (e.g., OCSD BAD wants to subscribe to threatening tweets from both Orange County and UCI) we can concatenate them using semicolons.

\begin{figure}[h]
\notsotiny
\vspace{-4mm}
\begin{lstlisting}[
           language=SQL,
           basicstyle=\ttfamily,
           showstringspaces=false,
           morekeywords={TYPE, DATASET, CREATE, FEED, WITH},
           commentstyle=\color{gray}
        ]
CREATE FEED A_SAMPLE_BAD_FEED_ON_ISLAND_B WITH { 
  "adapter-name" : "http_adapter", 
  "address-type" : "IP",
  "format" : "ADM", 
  "addresses" : "ISLAND_B_FEED_HOST:ISLAND_B_FEED_PORT",
  "type-name" : "INCOMING_DATA_TYPE", 
  "bad-host" : "ISLAND_A_HOST",
  "bad-channel" : "ISLAND_A_CHANNEL_NAME", 
  "bad-channel-parameters": "PARAM_1-1,PARAM_1-2;PARAM2-1,PARAM_2-2",
  "bad-dataverse": "ISLAND_A_CHANNEL_DATAVERSE" };
\end{lstlisting}
\vspace{-4mm}
\caption{Creating a BAD feed on island B}
\vspace{-2mm}
\label{ddl:create_bad_feed}
\end{figure}

The bridge's information is persisted in the BAD system's metadata with a feed's configuration when the feed is created. When a user starts a BAD feed on a local BAD system (island \textit{B}), it registers a broker on the specified remote BAD system (island \textit{A}) using island B's feed endpoint and subscribes to island A's channel using the provided parameters automatically. When a user stops the BAD feed, island \textit{B} unsubscribes from island A's channel and then removes the broker from island \textit{A}. We tie the start and stop events of a data feed on the local BAD system (island B) to the subscribe and unsubscribe actions on the remote BAD system (island A), so when the feed is not running, the remote BAD system will not need to compute and deliver data to this BAD feed.

\section{A Prototype of BAD Islands}
\label{sec:a_prototype_of_bad_islands}
We now describe a complete prototype of BAD islands that supports the use cases described in Section~\ref{sec:three_islands_example}. We show how to create and connect three BAD islands (the BAD trinity) using declarative statements and show how data flows between these different islands. The BAD system organizes data and other entities under \textit{dataverses} (similar to databases in an RDBMS). To differentiate organizations, we use different dataverses for different organizations (using the \textbf{USE} statement).

\subsection{BAD@DHS}
DHS BAD intakes tweets from external data sources. 
We can create a TweetFeed like the one in Figure~\ref{ddl:ing_feed} and configure it as dynamic to enrich the incoming tweets with additional information needed using UDFs. We first enrich an incoming tweet with the tweet's user's weapon registration records (if any). To hold the weapon registration records of sensitive tweet users, we create a data type \textit{WeaponRegistration} and a dataset \textit{WeaponRegistrations}, as shown in Figure~\ref{ddl:dhs_weapon_ds}. (A user may have multiple weapons.)

\begin{figure}[h]
\notsotiny
\vspace{-3mm}
\begin{lstlisting}[
           language=SQL,
           basicstyle=\ttfamily,
           showstringspaces=false,
           morekeywords={TYPE, DATASET, CREATE, FEED, WITH},
           commentstyle=\color{gray}
        ]
USE dhs;
CREATE TYPE WeaponRegistration AS 
  { wrid: uuid, uid: bigint, weapon_name: string };
CREATE DATASET WeaponRegistrations(WeaponRegistration) 
  PRIMARY KEY wrid AUTOGENERATED;
\end{lstlisting}
\vspace{-3mm}
\caption{Data type and dataset definition for weapon registration information}
\label{ddl:dhs_weapon_ds}
\end{figure}

Second, we create a Java UDF to detect the threatening rating of a tweet's text using a list of threatening words, as shown in Figure~\ref{udf:tweet_threatening_rating}. In this UDF, we load an external list of threatening words and we use the number of threatening words in the given text as its threatening rating. 

\begin{figure}[h]
\tiny
\vspace{-4mm}
\begin{lstlisting}[
           language=Java,
           basicstyle=\ttfamily,
           showstringspaces=false,
           commentstyle=\color{gray}
        ]
...
@Override
public void evaluate(IFunctionHelper functionHelper) throws Exception {
    JString input = (JString) functionHelper.getArgument(0);
    JInt output = (JInt) functionHelper.getResultObject();
    String tweetText = input.getValue();
    int threateningRating = 0;
    String[] words = tweetText.split(" ");
    for (String word : words) {
        // The threateningWordList is initialized with a file when function starts
        if (threateningWordList.contains(word.replaceAll("[,.]", ""))) {
            threateningRating++;
        }
    }
    output.setValue(threateningRating);
    functionHelper.setResult(output);
}
...
\end{lstlisting}
\vspace{-4mm}
\caption{A Java UDF for determining the threatening rating}
\vspace{-2mm}
\label{udf:tweet_threatening_rating}
\end{figure}

To add the desired set of enrichments to incoming tweets, we can create a SQL++ UDF \textit{EnrichTweet} and attach it to the TweetFeed when connecting to the Tweets dataset, as shown in Figure~\ref{ddl:dhs_tweet_udf}. In this UDF, we also transform the epoch time of a tweet's ``created\_at'' attribute into a datetime attribute ``timestamp'' and we create a point attribute ``location'' using the array of coordinates. These ADM attributes can be useful, as they do not need to be constructed in computations like spatial joins every time. Here we use the Java UDF defined in Figure~\ref{udf:tweet_threatening_rating} to extract the threatening rating of the tweet's text and attach it as a ``threatening\_rating'' attribute. We use a sub-query to look for the weapon registration information of the tweet's user and nest the registered weapons into a ``user\_registered\_weapon'' attribute. These new attributes are merged into the tweet and will be persisted for producing notifications.

\begin{figure}[h!]
\notsotiny
\vspace{-4mm}
\begin{lstlisting}[
           language=SQL,
           basicstyle=\ttfamily,
           showstringspaces=false,
           commentstyle=\color{gray}
        ]
USE dhs;
CREATE FUNCTION EnrichTweet(tweet) {
  object_merge(tweet, {
    "timestamp" : datetime_from_unix_time_in_ms(tweet.created_at), 
    "location" : 
       create_point(tweet.coordinates[0],tweet.coordinates[1]),
    "threatening_rating" : threateningRating(tweet.text),
    "user_registered_weapon": (SELECT VALUE w.weapon_name 
       FROM WeaponRegistrations w WHERE w.uid = tweet.uid)})
};
CONNECT FEED TweetFeed to DATASET Tweets APPLY FUNCTION EnrichTweet;
START FEED TweetFeed;
\end{lstlisting}
\vspace{-4mm}
\caption{Enriching tweets with additional information}
\vspace{-3mm}
\label{ddl:dhs_tweet_udf}
\end{figure}

\begin{figure}[h]
    \centering
    \includegraphics[width=0.45\textwidth]{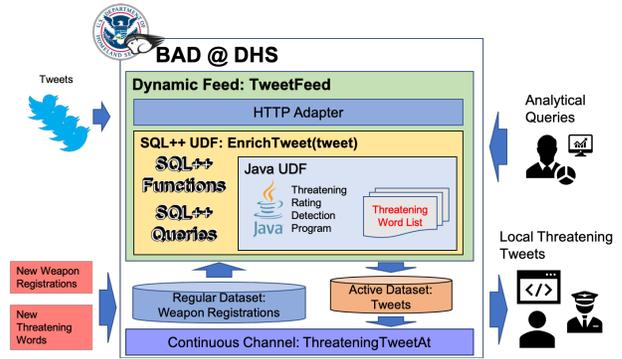}
    \caption{The internal details of the DHS BAD system}
    \vspace{-5mm}
    \label{fig:dhs_bad_overview}
\end{figure}

With these enriched threatening tweets, we can serve threatening tweets from areas by creating the continuous data channel ``ThreateningTweetsAt'' shown in Figure~\ref{ddl:dhs_channel}. To put everything together, a detailed overview of the entire DHS BAD system is shown in Figure~\ref{fig:dhs_bad_overview}.

\begin{figure}[h!]
\notsotiny
\vspace{-4mm}
\begin{lstlisting}[
           language=SQL,
           basicstyle=\ttfamily,
           showstringspaces=false,
           commentstyle=\color{gray}
        ]
USE dhs;
CREATE CONTINUOUS PUSH CHANNEL ThreateningTweetsAt(area_name) 
 PERIOD duration("PERIOD_DURATION") {
  SELECT t.area_name, t.text, t.location, t.threatening_rating,
    t.user_registered_weapon FROM Tweets t 
  WHERE t.area_name = area_name 
    AND t.threatening_rating > 0 AND is_new(t) };
\end{lstlisting}
\vspace{-4mm}
\caption{Definition of the ThreateningTweetsAt channel}
\vspace{-3mm}
\label{ddl:dhs_channel}
\end{figure}

\subsection{BAD@OCSD}
OCSD BAD in this prototype receives threatening tweets not only from Orange County but also from UCI to demonstrate how a BAD feed can connect to a channel with two sets of parameters. OCSD BAD notifies in-field officers about nearby threatening tweets that are close to important local events. To persist event information in OCSD BAD, we can create a data type ``Event'' and a dataset ``Events'' as shown in Figure~\ref{ddl:ocsd_event_ds}.

\begin{figure}[h!]
\notsotiny
\vspace{-4mm}
\begin{lstlisting}[
           language=SQL,
           basicstyle=\ttfamily,
           showstringspaces=false,
           commentstyle=\color{gray}
        ]
USE ocsd;
CREATE TYPE Event AS { eid: uuid, name: string, location: point, 
  event_duration: duration, radius_km: double };
CREATE DATASET Events(Event) PRIMARY KEY eid;
\end{lstlisting}
\vspace{-4mm}
\caption{Data type and dataset definition for events}
\vspace{-2mm}
\label{ddl:ocsd_event_ds}
\end{figure}

To store threatening tweets coming from DHS BAD, we create a data type \textit{LocalThreateningTweet} and an active dataset \textit{LocalThreateningTweets} in Figure~\ref{ddl:ocsd_threatening_tweet_ds}. (We use an active dataset for threatening tweets to ensure continuous query semantics in later local channel computations.)
We create a BAD feed in Figure~\ref{ddl:ocsd_threatening_tweet_feed} to obtain local threatening tweets from DHS. This BAD feed subscribes to the DHS threateningTweetsAt channel with parameters ``OC'' and ``UCI'', which correspond to two separate subscriptions in DHS BAD. Since there is no further data enrichment during ingestion, we use a static data feed here and connect it to LocalThreateningTweets directly.

\begin{figure}[h!]
\notsotiny
\vspace{-4mm}
\begin{lstlisting}[
           language=SQL,
           basicstyle=\ttfamily,
           showstringspaces=false,
           morekeywords={TYPE, DATASET, CREATE, FEED, WITH},
           commentstyle=\color{gray}
        ]
CREATE TYPE LocalThreateningTweet AS 
  { channelExecutionEpochTime: bigint, 
    dataverseName: string, channelName: string };
CREATE ACTIVE DATASET LocalThreateningTweets(LocalThreateningTweet) 
 PRIMARY KEY channelExecutionEpochTime;
\end{lstlisting}
\vspace{-4mm}
\caption{Data type and dataset definition for local threatening tweets at Orange County}
\vspace{-2mm}
\label{ddl:ocsd_threatening_tweet_ds}
\end{figure}

\vspace{-3mm}

\begin{figure}[h!]
\notsotiny
\vspace{-3mm}
\begin{lstlisting}[
           language=SQL,
           basicstyle=\ttfamily,
           showstringspaces=false,
           morekeywords={TYPE, DATASET, CREATE, FEED, WITH},
           commentstyle=\color{gray}
        ]
USE ocsd;
CREATE FEED LocalThreateningTweetFeed WITH {
  "adapter-name" : "http_adapter", 
  "addresses" : "OCSD_HOST:10013",
  "address-type" : "IP", 
  "type-name" : "LocalThreateningTweet", 
  "format" : "adm", 
  "bad-host" : "DHS_HOST", 
  "bad-channel" : "ThreateningTweetsAt",
  "bad-channel-parameters": "\"OC\";\"UCI\"", 
  "bad-dataverse": "dhs", 
  "dynamic": false };
CONNECT FEED LocalThreateningTweetFeed 
  TO DATASET LocalThreateningTweets;
START FEED LocalThreateningTweetFeed;
\end{lstlisting}
\vspace{-2mm}
\caption{Definition, connect and start feed statements for LocalThreateningTweetFeed}
\vspace{-2mm}
\label{ddl:ocsd_threatening_tweet_feed}
\end{figure}

In-field officers from OCSD also continuously send their location updates to the OCSD BAD system so that OCSD can notify the officers about nearby threatening tweets based on their current location.
We can use the data type, dataset, and feed described in Figure~\ref{ddl:channel} for intaking and persisting the location updates. As there is no further enrichment for location updates, the LocationFeed can be static as well.

With the local threatening tweets, event information, and officers' locations, we can now create a continuous channel for in-field officers to subscribe to nearby threatening tweets close to local events (a.k.a. threatening events), as shown in Figure~\ref{ddl:ocsd_channel}. The notifications from DHS contain threatening tweets as an array in the ``results'' attribute, so we use the UNNEST operation to access each independent threatening tweet. We calculate the distance between the officer and the tweet, the event and the tweet, and the officer and the event. If the officer is near a threatening tweet and the threatening tweet is near an event, we send a notification to the officer. The notification contains the tweet's content, the event information, the distance between the officer and the tweet, and the distance between the officer and the event in the notification to help the officer take further actions. A detailed overview of the OCSD BAD system is shown in Figure~\ref{fig:ocsd_bad_overview}.

\begin{figure}[h!]
\notsotiny
\vspace{-4mm}
\begin{lstlisting}[
           language=SQL,
           basicstyle=\ttfamily,
           showstringspaces=false,
           morekeywords={TYPE, DATASET, CREATE, FEED, WITH},
           commentstyle=\color{gray}
        ]
USE ocsd;
CREATE CONTINUOUS PUSH CHANNEL ThreateningEventsNear(oid) 
 PERIOD duration("PERIOD_DURATION") {
  FROM LocalThreateningTweets tn, OfficerLocations o, Events e
  UNNEST tn.results threatening_tweet
  LET tweet_loc = threatening_tweet.result.location,
  officer_tweet_dist = spatial_distance(o.location, tweet_loc),
  event_tweet_dist = spatial_distance(e.location, tweet_loc),
  officer_event_dist = spatial_distance(o.location, e.location)
    WHERE is_new(tn) AND oid = o.oid AND officer_tweet_dist < 0.1 
      AND event_tweet_dist < e.radius_km / 100
  SELECT oid, threatening_tweet.result tweet_content, e event_info, 
    officer_tweet_dist * 100 as tweet_distance_km, 
    officer_event_dist * 100 as event_distance_km
};
\end{lstlisting}
\vspace{-3mm}
\caption{Definition of the ThreateningEventsNear channel}
\vspace{-3mm}
\label{ddl:ocsd_channel}
\end{figure}

\vspace{-1.5mm}

\begin{figure}[h!]
    \centering
    \includegraphics[width=0.45\textwidth]{chapter4/ocsd_bad_overview.pdf}
\vspace{-2mm}
    \caption{The internal details of the OCSD BAD system}
\vspace{-3mm}
    \label{fig:ocsd_bad_overview}
\end{figure}

\subsection{BAD@UCI}
UCI BAD receives threatening tweets posted at UCI and checks whether a threatening tweet is near an on-campus building. If so, it creates a notification about the threatening tweet together with the nearby security stations' information. 
Like OCSD BAD, to persist threatening tweets at UCI, we need to create a data type \textit{LocalThreateningTweet} and a dataset \textit{LocalThreateningTweet} on UCI BAD. To receive threatening tweets at UCI from DHS, we need to create a BAD feed, like Figure~\ref{ddl:ocsd_threatening_tweet_feed}, connected to the ThreateningTweetsAt channel but using the parameter ``UCI''.

To provide more information for UCI BAD's subscribers, we store on-campus buildings, for checking whether there is a threatening tweet nearby, and security stations, for students to seek for help from, in UCI BAD. In Figure~\ref{ddl:uci_other_info}, we create the data types and datasets for them respectively.

\begin{figure}[h]
\notsotiny
\begin{lstlisting}[
           language=SQL,
           basicstyle=\ttfamily,
           showstringspaces=false,
           morekeywords={TYPE, DATASET, CREATE, FEED, WITH},
           commentstyle=\color{gray}
        ]
USE uci;
CREATE TYPE Building AS { bid: uuid, name: string };
CREATE TYPE SecurityStation AS { sid: bigint, location: point };
CREATE DATASET Buildings(Building) PRIMARY KEY bid AUTOGENERATED;
CREATE DATASET SecurityStations(SecurityStation) PRIMARY KEY sid;
\end{lstlisting}
\vspace{-2mm}
\caption{Data type and dataset definition of buildings and security stations}
\vspace{-5mm}
\label{ddl:uci_other_info}
\end{figure}

\begin{figure}[h!]
    \centering
    \includegraphics[width=0.45\textwidth]{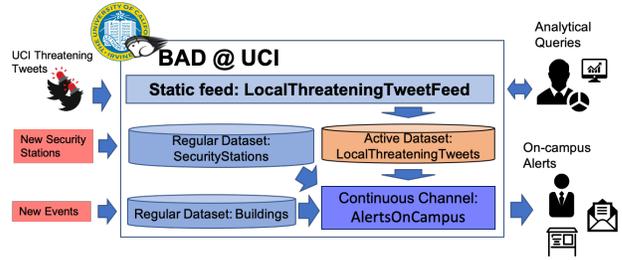}
\vspace{-3mm}
    \caption{The internal details of the UCI BAD system}
\vspace{-3mm}
    \label{fig:uci_bad_overview}
\end{figure}

With the local threatening tweets, on-campus building information, and security station information, we can create a continuous channel called ``AlertsOnCampus'' to provide on-campus alerts about threatening tweets near buildings with security stations' information attached using the statement shown in Figure~\ref{ddl:uci_channel}. Like the ThreateningEventsNear channel in OCSD BAD, we first UNNEST threatening tweets from the incoming notifications. Then, we check whether a threatening tweet is posted at an on-campus building. If so, we attach the security station information to the threatening tweet, with stations ordered by their distances to the tweet's location, and generate an alert. A detailed overview of the UCI BAD system is shown in Figure~\ref{fig:uci_bad_overview}.

\begin{figure}[h!]
\notsotiny
\vspace{-4mm}
\begin{lstlisting}[
           language=SQL,
           basicstyle=\ttfamily,
           showstringspaces=false,
           morekeywords={TYPE, DATASET, CREATE, FEED, WITH},
           commentstyle=\color{gray}
        ]
USE uci;
CREATE CONTINUOUS PUSH CHANNEL AlertsOnCampus() 
 PERIOD duration("PERIOD_DURATION") {
  FROM LocalThreateningTweets tn, Buildings b
  UNNEST tn.results threatening_tweet
  LET tweet_loc = threatening_tweet.result.location,
    station_dist = (FROM SecurityStations s
      LET dist = spatial_distance(tweet_loc, s.location)
      SELECT s stationInfo, dist * 100 dist_km ORDER BY dist)
  WHERE is_new(tn) AND spatial_intersect(tweet_loc, b.area)
  SELECT threateningTweet.result tweet_content, 
    b building_info, station_dist
};
\end{lstlisting}
\vspace{-2mm}
\caption{Definition of the AlertsOnCampus channel}
\vspace{-3mm}
\label{ddl:uci_channel}
\end{figure}

\vspace{-1.5mm}

\subsection{The Trip of A Threatening Tweet}
\begin{figure}[t!]
    \includegraphics[width=0.487\textwidth]{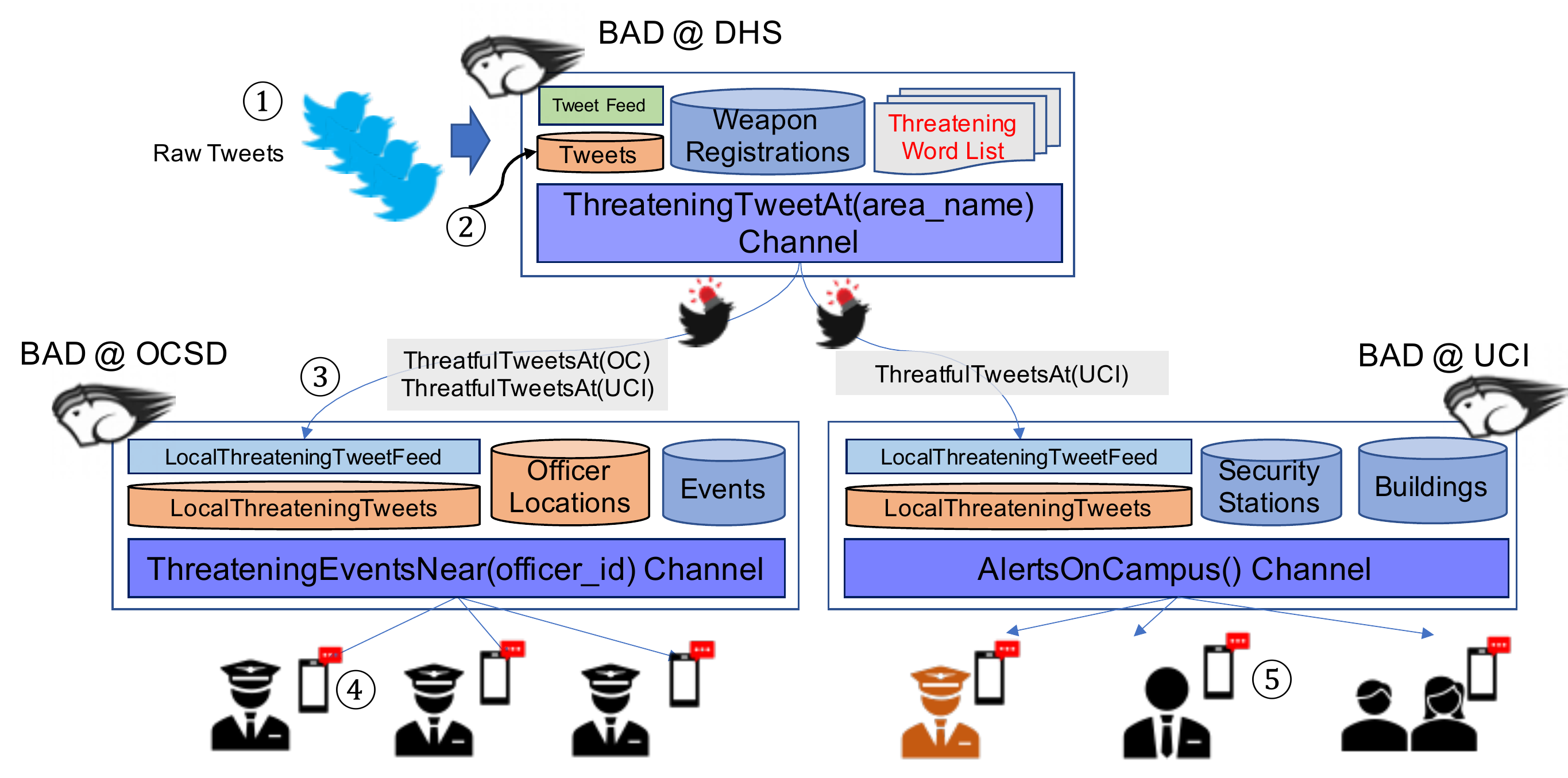}
    \caption{An overview of BAD islands}
    \vspace{-3.5mm}
    \label{fig:bad_islands_overview}
\end{figure}

In order to illustrate how BAD islands interact with BAD bridges, we pick a sample tweet and show how it flows through the three islands and their bridges and produces notifications with local information for the subscribers on each island.
An overview of our three-island prototype is shown in Figure~\ref{fig:bad_islands_overview}. The circled numerical labels in the figure will be used later for illustrating the data content at different stages of the workflow.

We will use the raw tweet in Figure~\ref{json:sample_tweet} (labeled 1 in Figure~\ref{fig:bad_islands_overview}) as the example. This tweet is posted at UCI, and it contains the tweet's geolocation as a JSON array of coordinates and the epoch timestamp of when the tweet was created as a JSON number. This raw tweet is ingested by the TweetFeed defined in Figure~\ref{ddl:ing_feed} and then enriched by the UDF defined in Figure~\ref{ddl:dhs_tweet_udf}. After that, the enriched tweet is persisted in the Tweets dataset as shown in Figure~\ref{json:enriched_tweet} (labeled 2 in Figure~\ref{fig:bad_islands_overview}). Enriched tweets contain a threatening rating detected by the Java UDF, an array of registered weapons for the tweet's user obtained by looking in the WeaponRegistrations dataset using the ``uid'' attribute, a timestamp as a datetime attribute, and a location as a point attribute.

\begin{figure}[t!]
\notsotiny
\vspace{-1mm}
\begin{lstlisting}[
           language=json,
           basicstyle=\ttfamily,
           showstringspaces=false,
           commentstyle=\color{gray}
        ]
{
  "tid": 1593142018123,
  "uid": 73,
  "area_name": "UCI",
  "text": "Saul Goodman builds SKS, and Todd Alquist fires AK47, 
           but Skyler White sells Cabbage.",
  "coordinates": [ 33.64921228736088, -117.84181977473024 ],
  "created_at": 1593142018123
}
\end{lstlisting}
\vspace{-4mm}
\caption{A sample raw threatening tweet}
\vspace{-5mm}
\label{json:sample_tweet}
\end{figure}

\begin{figure}[t!]
\notsotiny
\begin{lstlisting}[
           language=json,
           basicstyle=\ttfamily,
           showstringspaces=false,
           commentstyle=\color{gray}
        ]
{
  "tid": 1593142018123,
  "uid": 73,
  "area_name": "UCI",
  "text": "Saul Goodman builds SKS, and Todd Alquist fires AK47, 
           but Skyler White sells Cabbage.",
  "coordinates": [ 33.64921228736088, -117.84181977473024 ],
  "created_at": 1593142018123,
  "threatening_rating": 2,
  "user_registered_weapon": [ "AR10" , "AK47", "GLOCK21" ],
  "timestamp": datetime("2020-06-26T03:26:58.123Z"),
  "location": point("33.64921228736088,-117.84181977473024")
}
\end{lstlisting}
\vspace{-4mm}
\caption{The enriched threatening tweet}
\vspace{-5mm}
\label{json:enriched_tweet}
\end{figure}

Since both the OCSD and UCI BAD systems subscribe to threatening tweets at UCI, they will each receive a notification from DHS BAD about this threatening tweet. Figure~\ref{json:dhs_notification} shows the notification sent to OCSD BAD (labeled 3 in Figure~\ref{fig:bad_islands_overview}). If there was also a threatening tweet posted in Orange County at the same time, the ``results'' array would include that tweet but with a different subscription ID, as OCSD BAD has two subscriptions to the ThreateningTweetAt channel with parameters ``OC'' and ``UCI'', respectively. Since UCI BAD also subscribes to the channel, but with a different subscription on another broker (pointed to UCI's BAD feed), the notification for UCI BAD will be produced and sent separately.

\begin{figure}[t]
\notsotiny
\begin{lstlisting}[
           language=json,
           basicstyle=\ttfamily,
           showstringspaces=false,
           commentstyle=\color{gray}
        ]
{
  "dataverseName": "dhs",
  "channelName": "ThreateningTweetsAt",
  "channelExecutionEpochTime": 1593142019521,
  "results": [
    {
      "result": {
        "text": "Saul Goodman builds SKS, and Todd Alquist fires 
                AK47, but Skyler White sells Cabbage.",
        "area_name": "UCI",
        "location": point("33.64921228736088,-117.84181977473024"),
        "threatening_rating": 2,
        "user_registered_weapon": [ "AR10" , "AK47", "GLOCK21" ]
      },
      "channelExecutionTime": datetime("2020-06-26T03:26:59.521Z"),
      "subscriptionId": uuid("82e61d25-f7ad-0632-3b9a-9c26e681ad84"),
      "deliveryTime": datetime("2020-06-26T03:26:59.522Z")
    }
  ]
}
\end{lstlisting}
\vspace{-4mm}
\caption{The generated threatening tweet notification from DHS}
\vspace{-5mm}
\label{json:dhs_notification}
\end{figure}

In the OCSD BAD ThreateningEventsNear channel, threatening tweets are combined with local event information and officer location information to produce the nearby threatening event notifications for in-field officers. 
There is one local event ``OC Marathon'' near the threatening tweet in Figure~\ref{json:enriched_tweet}, and there is an in-field officer 0 nearby, so OCSD BAD produces one notification about the tweet and the event for this officer. Figure~\ref{json:ocsd_threatening_event} shows this threatening event notification (labeled 4 in Figure~\ref{fig:bad_islands_overview}). It contains the event information as the ``event\_info'' attribute, the threatening tweet's information as the ``tweet\_content'' attribute, and the distances from the officer 0 to the tweet and to the event as the ``event\_distance\_km'' and ``tweet\_distance\_km'' attributes respectively. 

\begin{figure}[t]
\notsotiny
\begin{lstlisting}[
           language=json,
           basicstyle=\ttfamily,
           showstringspaces=false,
           commentstyle=\color{gray}
        ]
{
  "dataverseName": "ocsd",
  "channelName": "ThreateningEventsNear",
  "channelExecutionEpochTime": 1593142020436,
  "results": [
    {
      "result": {
        "event_info": {
          "eid": uuid("82e61d25-4cad-0632-3d8d-148e71cb50bf"),
          "name": "OC Marathon",
          "location": 
             point("33.66100302712824, -117.83950620703125"),
          "event_duration": duration("PT10S"),
          "radius_km": 3.57746886883645
        },
        "tweet_distance_km": 4.854786471222485,
        "event_distance_km": 5.6839370484947755,
        "oid": 0,
        "tweet_content": {
          "text": "Saul Goodman builds SKS, and Todd Alquist fires 
                    AK47, but Skyler White sells Cabbage.",
          "area_name": "UCI",
          "location": point("33.64921228736088,-117.84181977473024"),
          "threatening_rating": 2,
          "user_registered_weapon": [ "AR10" , "AK47", "GLOCK21" ]
        }
      },
      "channelExecutionTime": datetime("2020-06-26T03:27:00.436Z"),
      "subscriptionId": uuid("82e61d25-47ad-0632-3e5c-22b3cb7d7df4"),
      "deliveryTime": datetime("2020-06-26T03:27:00.437Z")
    }
  ]
}
\end{lstlisting}
\vspace{-4mm}
\caption{The generated threatening event notification from OCSD}
\vspace{-5mm}
\label{json:ocsd_threatening_event}
\end{figure}

In the UCI BAD AlertsOnCampus channel, threatening tweets are combined with on-campus building information and security station information to produce alerts. The threatening tweet in Figure~\ref{json:enriched_tweet} is near the building ``Student Center'', so UCI BAD produces a notification to alert people around this building as shown in Figure~\ref{json:uci_alert} (labeled 5 in Figure~\ref{fig:bad_islands_overview}). The building information is attached to the notification.
There are two security stations nearby, so the system attaches their information with their distances, ordered by their distances to the threatening tweet. Everyone subscribing to the AlertsOnCampus channel will receive this notification.

\begin{figure}[h]
\notsotiny
\begin{lstlisting}[
           language=json,
           basicstyle=\ttfamily,
           showstringspaces=false,
           commentstyle=\color{gray}
        ]
{
  "dataverseName": "uci",
  "channelName": "AlertsOnCampus",
  "channelExecutionEpochTime": 1593142024344,
  "results": [
    {
      "result": {
        "buildingInfo": {
          "bid": uuid("82e61d25-43ad-0632-45d0-0ba5366832d9"),
          "name": "Student Center",
          "area": rectangle("33.64811430275051, -117.84332027249145
                    33.649382536086605,-117.84153928570557")
        },
        "stationDist": [
          {
            "stationInfo": {
              "sid": 1,
              "location": 
                 point("33.64792551859947, -117.84013290702327"),
              "name": "Station # 1"
            },
            "dist_km": 0.21216259109805177
          },
          {
            "stationInfo": {
              "sid": 0,
              "location": 
                 point("33.646866723393266, -117.84170161534618"),
              "name": "Station # 0"
            },
            "dist_km": 0.23485382616041114
          }
        ],
        "tweetContent": {
          "text": "Saul Goodman builds SKS, and Todd Alquist fires 
                    AK47, but Skyler White sells Cabbage.",
          "area_name": "UCI",
          "location": point("33.64921228736088,-117.84181977473024"),
          "threatening_rating": 2,
          "user_registered_weapon": [ "AR10" , "AK47", "GLOCK21" ]
        }
      },
      "channelExecutionTime": datetime("2020-06-26T03:27:04.344Z"),
      "subscriptionId": uuid("82e61d25-0ead-0632-4717-e17b6a912fa6"),
      "deliveryTime": datetime("2020-06-26T03:27:04.345Z")
    }
  ]
}
\end{lstlisting}
\vspace{-5mm}
\caption{The generated on-campus alert from UCI}
\vspace{-2mm}
\label{json:uci_alert}
\end{figure}

\section{BAD Islands Tour and Evaluation}
\label{sec:bad_islands_tour}
\begin{figure}[h]
    \centering
    \includegraphics[width=0.48\textwidth]{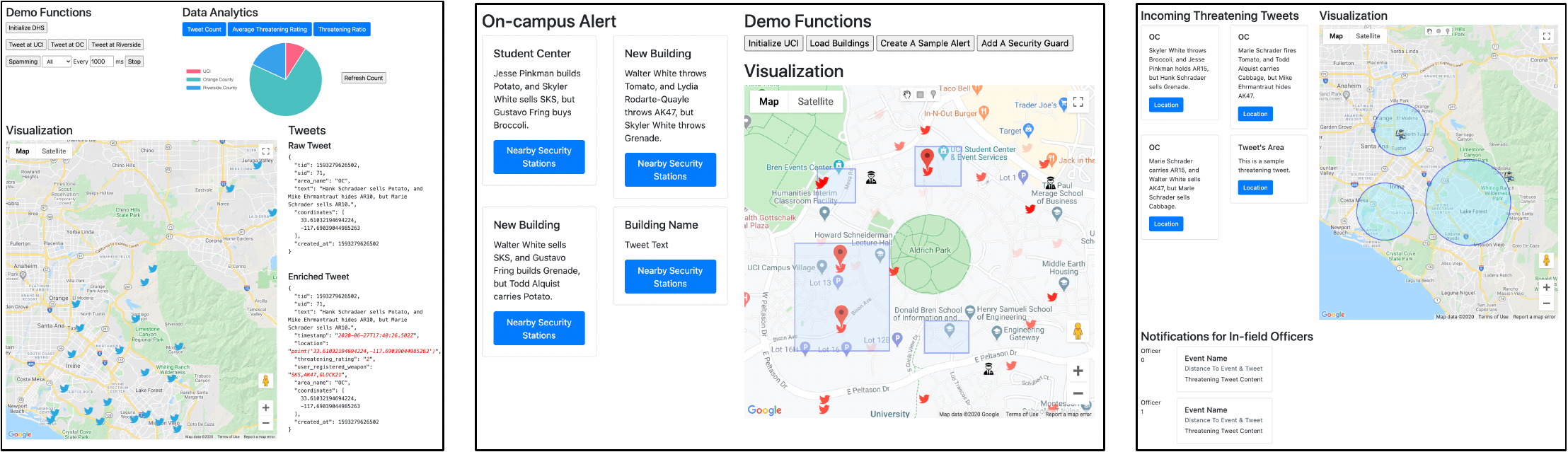}
\vspace{-2mm}
    \caption{An overview of BAD islands dashboards}
    \vspace{-7mm}
    \label{fig:demo_overview}
\end{figure}

To illustrate how BAD applications can be built with BAD islands and to visualize the process of data flowing through multiple systems and becoming notifications for subscribers, we have created three dashboards for each organization based on our prototype, as shown in Figure~\ref{fig:demo_overview}.

\begin{figure}[h]
    \centering
    \includegraphics[width=0.48\textwidth]{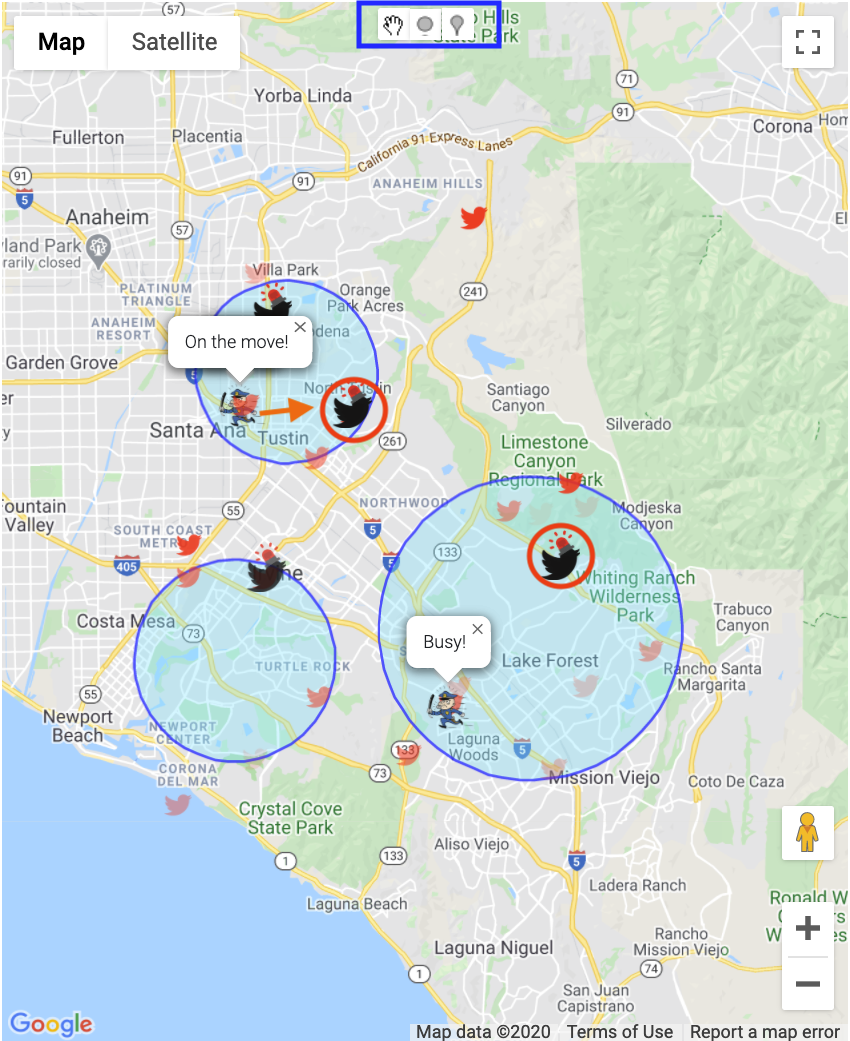}
    \vspace{-2mm}
    \caption{Visualization panel of OCSD dashboard}
    \vspace{-5mm}
    \label{fig:ocsd_dash_visualization}
\end{figure}

Due to space limits, instead of describing the features on each dashboard in detail, we will focus on the Visualization Panel of the OCSD Dashboard, shown in Figure~\ref{fig:ocsd_dash_visualization}, to illustrate how threatening tweets go from DHS BAD to OCSD BAD and how OCSD BAD combines threatening tweets with other local information for its subscribers.

The Visualization Panel contains a map that shows the incoming threatening tweets, local events, produced threatening events, and in-field officers' movements. The map contains a control bar at the top (highlighted in a blue box) so dashboard users can navigate the map, add a new event, and add a new in-field officer. A new event can be added by drawing a circle on the map indicating the event's area. An officer can be added by dropping an officer icon on a preferred location on the map. Information about the created events and officers is updated in the underlying OCSD BAD system accordingly. An added officer moves around the map randomly and continuously sends its current location to the OCSD BAD system. One can change an officer's location by dragging the officer's icon to a new place on the map.

We use a red tweet icon to mark the threatening tweets received from DHS BAD and a black tweet icon for the threatening events detected by OCSD BAD. When an in-field officer receives a threatening event notification (as highlighted in the red circles in the figure), the officer randomly decides whether to go to the threatening event's location for further investigation or to stay at his or her current location. The officer's decision pops up as a small information window, as shown in the figure. If the officer decides to go, he or she moves gradually towards the tweet's location, as the upper officer does in the figure.

In addition to the dashboards, we also conducted a simple experiment to measure the tweet propagation delays in our prototype system, starting from the posting of new tweets to the receipt of the localized notifications by subscribers on each island. We deployed the prototype on a three-node cluster, one node per island, where each node had a Dual-Core AMD Opteron Processor 2212 2.0 GHz, 8 GB of RAM, and a 900 GB hard disk drive. We used the statements described in Section~\ref{sec:a_prototype_of_bad_islands} to configure the nodes.

\begin{figure}[h]
\vspace{-2mm}
    \centering
    \includegraphics[width=0.48\textwidth]{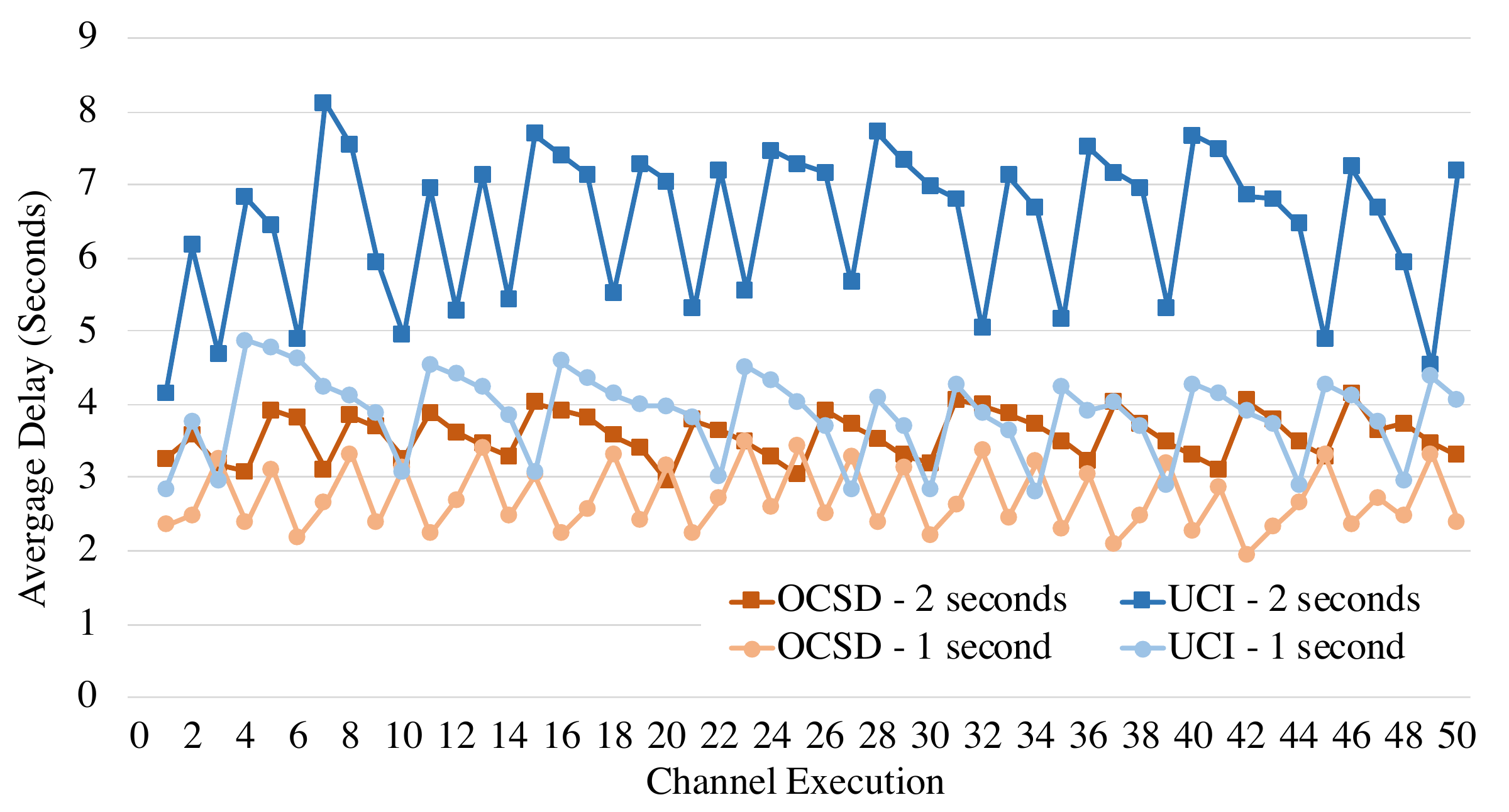}
\vspace{-2mm}
    \caption{Delays of threatening tweets for OCSD and UCI BAD subscribers}
    \label{fig:delay}
\end{figure}

The information propagation times for BAD islands depend on the complexity of the computations in the pipeline (data enrichment and channel computation) and on the specified channel period durations. In our experiments, we used the same channel period for all three channels, testing two different channel periods (1s and 2s).
Since channels execute once per each channel period, for each channel execution, we measured the average delay for threatening tweets delivered to subscribers in this channel execution. We let all channels complete 50 executions and kept track of the average delays throughout the process.
On DHS BAD, tweets were set to arrive at 10 tweets per second, and half of the tweets contained at least one threatening word. 
On OCSD BAD, every threatening tweet had an event nearby. OCSD had 100 in-field officers constantly updating their locations and subscribing to nearby threatening events.
On UCI BAD, every threatening tweet was close to an on-campus building. UCI had 5 on-campus security stations and 100 subscriptions subscribing to on-campus alerts.
The delays are shown in Figure~\ref{fig:delay}. 

Clearly, the subscribers at OCSD and UCI are able to receive localized threatening tweets of interest in a timely manner, while the delays are relatively stable, especially for the 1s channel period. When the channel period is increased to 2s, the delays increase since the system batches more incoming tweets per channel execution; 
while this would increase the delay for subscribers, it would also increase the system scalability under higher loads.
UCI BAD in general has higher delays than OCSD BAD due to a more complex computation and more local information added to local threatening tweets.

\section{Conclusions}
\label{sec:bad_islands_summary}
In this work, we have focused on enabling users to declaratively create scalable data sharing services between different BAD systems. 
We looked at an example use case in which two local organizations (OCSD and UCI) would like to get data from a third organization (DHS) in order to provide BAD services to their subscribers.
We discussed several possible ways of supporting this use case and proposed using data feeds and data channels for bridging BAD systems.
We extended the BAD system with \textit{BAD brokers} to simplify data exchanges between channels and feeds and \textit{BAD feeds} to help users create bridges between different BAD systems.
We detailed a three-island prototype to show how BAD islands can be bridged together. We demonstrated how users can easily build such systems with declarative statements, and we used an example to show how data and events flow within the system.
We built a set of dashboards based on our prototype to concretely illustrate how BAD islands share data and support BAD applications with localized information, and we conducted an experiment to examine the delays in the prototype system.

\section*{Acknowledgment}
\vspace{-1.5mm}
This research was partially supported by NSF grants IIS-1447826, IIS-1447720, IIS-1838222, IIS-1838248, CNS-1924694 and CNS-1925610.





\renewcommand{\IEEEbibitemsep}{0pt plus 0.5pt}
\makeatletter
\IEEEtriggercmd{\reset@font\normalfont\fontsize{8pt}{8.8pt}\selectfont}
\makeatother
\IEEEtriggeratref{1}
\vspace{-1.5mm}
\bibliographystyle{./bibliography/IEEETran}
\bibliography{./bibliography/IEEEexample}

\end{document}